\documentclass{article}
\usepackage{spconf,amsmath,graphicx}
\usepackage{booktabs}

\usepackage{amsmath,epsfig,amssymb,amsthm,url,amsfonts}
\usepackage{algorithm}% http://ctan.org/pkg/algorithm
\usepackage{algpseudocode}% http://ctan.org/pkg/algorithmicx
\usepackage{multirow}
\usepackage{mdframed}
\usepackage{url}
\usepackage{enumitem}
\usepackage[makeroom]{cancel}
\usepackage{dblfloatfix}
\usepackage{array}
\usepackage{epstopdf}
\usepackage{float}
\usepackage{subfig}
\usepackage{breqn}
\usepackage{cite}
\usepackage{xcolor}
\usepackage{tabularx}
\usepackage[printonlyused,withpage]{acronym}
\usepackage{balance}

\theoremstyle{break}

\theoremstyle{definition}

\newcommand{\E}{\mathbb{E}}

\def\mbf#1{\mathbf{#1}}

\def\mbb#1{\mathbb{#1}}

\title{Robust Unsupervised Audio-visual Speech Enhancement\\ Using a Mixture of Variational Autoencoders}
\name{Mostafa Sadeghi and Xavier Alameda-Pineda, IEEE Senior Member\thanks{Xavier Alameda-Pineda acknowledges ANR and the IDEX for funding the ML3RI project. }}
\address{Inria Grenoble Rh\^{o}ne-Alpes, France}
\newcommand{\lk}{ \left\{ }
\newcommand{\rk}{ \right\} }

\newcommand{\Rbb}{{\mathbb{R}}}

\newcommand{\Hb}{{\bf H}}

\newcommand{\sbb}{{\bf s}}

\newcommand{\xb}{{\bf x}}

\newcommand{\zb}{{\bf z}}

\newcommand{\Wb}{{\bf W}}

\newcommand{\Ib}{{\bf I}}

\newcommand{\vb}{{\mathbf v}}
\newcommand{\Vb}{{\mathbf V}}

\newcommand{\Nc}{{\cal N}}

\newsavebox\mybox

\acrodef{SE}{speech enhancement}
\acrodef{STFT}{short-time Fourier transform}
\acrodef{PSD}{power spectral density}
\acrodef{NMF}{nonnegative matrix factorization}
\acrodef{AV}{audio-visual}
\acrodef{DNN}{deep neural network}
\acrodefplural{DNNs}{deep neural networks}
\acrodef{VAE}{variational auto-encoder}
\acrodefplural{VAEs}{variational auto-encoders}
\acrodef{CVAE}{conditional variational auto-encoder}
\acrodefplural{CVAEs}{conditional variational auto-encoders}
\acrodef{A-VAE}{audio VAE}
\acrodef{V-VAE}{visual VAE}
\acrodef{AV-CVAE}{audio-visual CVAE}
\acrodef{ROI}{region of interest}
\acrodef{MCMC}{Markov Chain Monte Carlo}
\acrodef{EM}{expectation-maximization}
\acrodef{MCEM}{Monte Carlo expectation-maximization}
\acrodef{TF}{time frequency}
\acrodef{ELBO}{evidence lower bound}
\acrodef{ROI}{region of interest}
\acrodef{LR}{Living Room}
\acrodef{SDR}{signal-to-distortion ratio}
\acrodef{PESQ}{perceptual evaluation of speech quality}
\acrodef{ASE}{audio speech enhancement}
\acrodef{VSE}{visual speech enhancement}
\acrodef{AVSE}{audio-visual speech enhancement}
\acrodef{SNR}{signal-to-noise ratio}
\acrodefplural{SNRs}{signal-to-noise ratios}
\acrodef{LSTM}{long short-term memory}

\begin{document}
%\ninept
%
\maketitle
\begin{abstract}
Recently, an audio-visual speech generative model based on variational autoencoder (VAE) has been proposed, which is combined with a \ac{NMF} model for noise variance to perform unsupervised speech enhancement. When visual data is clean, speech enhancement with audio-visual VAE shows a better performance than with audio-only VAE, which is trained on audio-only data. However, audio-visual VAE is not robust against noisy visual data, e.g., when for some video frames, speaker face is not frontal or lips region is occluded. In this paper, we propose a robust unsupervised audio-visual speech enhancement method based on a per-frame VAE mixture model. This mixture model consists of a trained audio-only VAE and a trained audio-visual VAE. The motivation is to skip noisy visual frames by switching to the audio-only VAE model. We present a variational expectation-maximization method to estimate the parameters of the model. Experiments show the promising performance of the proposed method.

\end{abstract}
\section{Introduction}
Speech enhancement -- or how to estimate clean speech from a noisy signal -- has attracted a lot of attention, both for single- and 
multi-channel audio recordings~\cite{VincVG18,wang2018supervised,loizou2007speech,xu2015regression}. Recently, generative models have been utilized for 
speech enhancement \cite{bando2018statistical,Leglaive_MLSP18,SekiguchiAPSIPA2018, Leglaive_ICASSP2019a,Leglaive_ICASSP2019b,PariDV19}. Specifically, some 
works proposed to use variational autoencoder (VAE) to model speech spectrogram, and then perform speech enhancement by considering an \ac{NMF} noise 
variance model~\cite{bando2018statistical,Leglaive_MLSP18}. This is done in an unsupervised way. A common characteristic of all these methods, which 
we refer to as \ac{A-VAE}, is the use of audio recordings only.

Audio-visual speech enhancement methods incorporate also the visual information (video frames) associated with the noisy speech, aiming to improve the 
quality of the enhanced speech signal \cite{girin2001audio,AfouCZ18,GabbSP18}. Using the video modality is well-motivated by the fact that lips movements provide information about what is being uttered. As an audio-visual extension of VAE-based methods of \cite{bando2018statistical,Leglaive_MLSP18}, an audio-visual VAE (AV-VAE) model has recently been proposed 
in~\cite{sadeghiLAGH19}, training a VAE model conditioned on visual features, e.g.,\ lips \ac{ROI}. For speech enhancement, AV-VAE has been shown 
to outperform A-VAE specially in high noise levels~\cite{sadeghiLAGH19}.

A critical problem with audio-visual methods is ``noisy visual data'', e.g., when speakers do not face the camera or the lips are occluded. Specifically, the AV-VAE based speech enhancement method presented in~\cite{sadeghiLAGH19} uses clean visual data to train the speech spectrogram prior. Therefore, it expects clean visual data as well in the test (enhancement) phase for unseen data. Otherwise, its performance could degrade below audio-only methods, as we will see later in this paper.

The present work aims to provide a solution to the above-mentioned problem. That is, to make AV-VAE speech enhancement robust against noisy visual data. 
To achieve this goal, we propose a VAE mixture model consisting of a trained A-VAE and a trained AV-VAE model. As said before, AV-VAE yields poor results in the presence of noisy visual data. However, the proposed mechanism would skip visual data whenever it is not reliable, and uses the A-VAE model instead. Importantly, the choice between A-VAE and AV-VAE is unsupervised and must be done for every frame at test time. We present a variational inference framework to tackle all these issues. Experimental results on clean as well as noisy visual frames are provided, demonstrating the effectiveness of our method.

The rest of this paper is organized as follows. Section~\ref{sec:review} briefly reviews the audio-only and audio-visual VAE-based speech spectrogram modeling 
methods. Section~\ref{sec:prop} presents the proposed mixture generative model. Sections~\ref{sec:inf} details the inference and speech 
reconstruction steps. Finally, experimental results are provided in Section~\ref{sec:exp}.

\section{VAE-based Speech Spectra Modeling}\label{sec:review}
\subsection{Audio-only VAE}
In this section, we briefly review the VAE generative speech model that was first proposed in \cite{bando2018statistical}. Let $s_{fn}$ denote the complex-valued speech \ac{STFT} coefficient at frequency index $f \in \{0,...,F-1\}$ and at frame index $n \in \{0,...,N-1\}$. At each \ac{TF} bin, we have the following probabilistic generative model, which will be referred to as \ac{A-VAE}:
\begin{align}
\label{decoder_VAE}
s_{fn} | \mathbf{z}_n &\sim \mathcal{N}_c(0, \sigma_{f}^a(\mathbf{z}_n)), \\
\label{prior_VAE}
\mathbf{z}_n &\sim \mathcal{N}(\mathbf{0}, \mathbf{I}), 
\end{align}
where $\mathbf{z}_n \in \mathbb{R}^L$, with $L \ll F$, is a latent random variable describing a speech generative process, 
$\mathcal{N}(\mathbf{0}, \mathbf{I})$ is a zero-mean multivariate Gaussian distribution with identity covariance matrix, and 
$\mathcal{N}_c(0, \sigma)$ is a
univariate complex proper Gaussian distribution with zero mean and variance $\sigma$. Let $\mathbf{s}_n\in \mathbb{C}^F$ be the vector whose components are the speech \ac{STFT} coefficients at frame $n$. The set of non-linear functions $\{\sigma_{f}^a: \mbb{R}^L \mapsto 
\mbb{R}_+\}_{f=0}^{F-1}$ are modeled as neural networks  sharing the input $\mbf{z}_n \in 
\mathbb{R}^L$. These parameters are estimated using variational inference by defining another neural network, called encoder (inference) network, which approximates the intractable posterior of $ \zb_n $ given $ \sbb_n $  \cite{KingW14, bando2018statistical, Leglaive_MLSP18}.

\subsection{Audio-visual VAE}
In the AV-VAE framework proposed in \cite{sadeghiLAGH19}, the following latent space model is considered, independently for all $l \in \{0,...,L-1\}$ 
and all TF bins $(f,n)$:
\begin{align}
\label{eq:model_cvae}
s_{fn} | \mathbf{z}_n, \vb_n &\sim \mathcal{N}_c(0, \sigma_{f}^{av}(\mathbf{z}_n, \vb_n)),\\
\label{eq:latent_cvae}
z_{ln} | \vb_n &\sim \mathcal{N}\left(\bar{\mu}_l(\vb_n), \bar{\sigma}_{l}(\vb_n)\right),
\end{align}
where $\vb_n  \in \mathbb{R}^M$ is an embedding for the image of the speaker lips at frame $ n $, and the non-linear functions $\{{\sigma}_f^{av}:\Rbb^L\times \Rbb^M \mapsto \mathbb{R}_+\}_{f=0}^{F-1}$ are modeled as a neural network taking $ \zb_n $ and $ \vb_n $ as input. Furthermore, the non-linear functions $\{\bar{\mu}_l:\Rbb^M \mapsto \mathbb{R}\}_{l=0}^{L-1}$ and $\{\bar{\sigma}_l:\Rbb^M \mapsto \mathbb{R}_+\}_{l=0}^{L-1}$, yielding $\zb_n$'s prior, are collectively modeled with a neural network which takes $ \vb_n $ as input. In a way similar to A-VAE, an encoder network, approximating the intractable posterior of 
$\mathbf{z}_n$ given $\mathbf{s}_n$ and $\mathbf{v}_n$, is defined and trained jointly with the decoder \eqref{eq:model_cvae} and prior \eqref{eq:latent_cvae} using a 
clean set of speech spectrogram frames and the corresponding clean visual data, i.e., images of lips region \cite{sadeghiLAGH19}. 

Now that both A-VAE and AV-VAE are trained 
for their specific tasks, in the following, we present the VAE mixture model to automatically select one of the two at inference time.

\section{VAE Mixture Model}\label{sec:prop}
To make the speech enhancement robust to noisy visual data, we propose an automatic selection mechanism between A-VAE and AV-VAE. Ideally, such a mechanism would 
allow to select the best-suited method at each frame: when visual information is clean, AV-VAE, and otherwise, A-VAE. We formalise this with a mixture model, named 
VAE mixture model (VAE-MM):
\begin{align}
p(s_{fn}|\zb_n,\vb_n,\alpha_n) &= \Big[\Nc_c(0, \sigma_f^a(\zb_n))\Big]^{\alpha_n}\times\\
&\phantom{{}=1}\Big[\Nc_c\left(0, \sigma_f^{av}(\zb_n,\vb_n)\right)\Big]^{1-\alpha_n},\nonumber\\
p(\zb_n|\vb_n,\alpha_n) &= \Big[\Nc(\boldsymbol{0}, \Ib)\Big]^{\alpha_n}\times\\
&\phantom{{}=1} \Big[\prod_l \Nc\left(\bar{\mu}_l(\vb_n), \bar{\sigma}_l(\vb_n)\right)\Big]^{1-\alpha_n}\nonumber\\
p(\alpha_n) &= \pi^{\alpha_n} \times (1-\pi)^{1-\alpha_n},
\end{align}
where, $ \alpha_n\in \lk 0,1 \rk $ is a latent variable specifying the component of the mixture model that is used by the $ n $-th frame, for 
both $s_{fn}$ and $\zb_n$. The prior distribution of $ \alpha_n $ is chosen as a Bernoulli distribution with parameter $ \pi $. 

\section{VAE-MM Inference \& Learning}\label{sec:inf}
The observed noisy microphone signal writes:
\begin{equation}
x_{fn} = s_{fn} + b_{fn},%~~~\forall(f,n) \in \mathbb{B},
\label{mix_model}
\end{equation}
for all \ac{TF} bins $ (f,n) $. Similarly as done in the previous works~\cite{bando2018statistical,Leglaive_MLSP18,SekiguchiAPSIPA2018, 
Leglaive_ICASSP2019a,Leglaive_ICASSP2019b,PariDV19}, we use an unsupervised \ac{NMF}-based Gaussian noise model that 
assumes independence across \ac{TF} bins:%. That is, independently for all TF bins $(f,n)$:
\begin{equation}
b_{fn} \sim \mathcal{N}_c\left(0, \left(\mathbf{W}_b\mathbf{H}_b\right)_{fn}\right),
\label{eq:noise_model}
\end{equation}
where $ \mathbf{W}_b\in\Rbb^{F\times K}$ is a nonnegative matrix of spectral power patterns and $\mathbf{H}_b\in\Rbb^{K\times N} $ is a nonnegative matrix of 
temporal activations, with $K$ such that $ K(F + N ) \ll F N $.

The set of parameters to be estimated is defined as $ \Theta = \lk 
\mathbf{W}_b,\mathbf{H}_b, \pi \rk $. We use a variational expectation-maximization (VEM) approach~\cite{bishop06} to estimate these parameters. To do so, the 
intractable posterior $ p(\sbb_n, \zb_n, \alpha_n|\xb_n) 
$ is approximated by a variational distribution factorizing as follows:
\begin{equation}
\label{eq:posterior}
r(\sbb_n, \zb_n, \alpha_n) = r(\sbb_n)\; r(\zb_n)\; r(\alpha_n).
\end{equation}
The variational factors in \eqref{eq:posterior} are then estimated by minimizing the Kullback-Leibler divergence between \eqref{eq:posterior} and the true 
posterior. The final update formulas for the variational distributions are given below and are used in an alternating optimization strategy. The details of the derivations are provided in a supporting document which is available online.\footnote{\url{https://team.inria.fr/perception/research/vae-mm-se/}}
\subsection{E-$ \sbb_n $ step}
The variational distribution of $ \sbb_n $ factorizes over $ f $. For each component, we obtain the following:
\begin{equation}
r(s_{fn}) = \Nc_c(m_{fn},\nu_{fn}),
\end{equation}
where
\begin{equation}
\label{eq:rs0}
\begin{cases}
m_{fn} & = \frac{\gamma_{fn}}{\gamma_{fn} + \left(\mathbf{W}_b\mathbf{H}_b\right)_{fn}}\cdot x_{fn} \\
\nu_{fn} & = \frac{\gamma_{fn}\cdot \left(\mathbf{W}_b\mathbf{H}_b\right)_{fn}}{\gamma_{fn} + \left(\mathbf{W}_b\mathbf{H}_b\right)_{fn}}
\end{cases},
\end{equation}
\begin{equation}
\gamma_{fn}^{-1}=\sum_{\alpha_n\in\lk 0,1\rk}r(\alpha_n)\cdot\eta_{fn}^{\alpha_n},
\label{eq:gamma}
\end{equation}
\begin{equation}
\label{eq:gamma_}
\eta_{fn}^{\alpha_n} = \E_{r(\zb_n)}\left[\frac{1}{\sigma_f^{\alpha_n}(\zb_n,\vb_n)}\right]\approx\frac{1}{D}\sum_{d=1}^{D} \frac{1}{\sigma_f^{\alpha_n}(\zb_{n}^{(d)},\vb_n)},
\end{equation}
and $\lk\zb_n^{(d)}\rk_{d=1}^D$ is a sequence sampled from $ r(\zb_n) $. Moreover, we have defined $ \sigma_f^{\alpha_n}(\zb_n,\vb_n) $ as follows:
\begin{equation}
\label{eq:rs1}
\sigma_f^{\alpha_n}(\zb_n,\vb_n)=\begin{cases}
\sigma_f^{a}(\zb_n) & \alpha_n = 1 \\
\sigma_f^{av}(\zb_n,\vb_n) & \alpha_n = 0
\end{cases}.
\end{equation}
\subsection{E-$ \zb_n $ step}
For $ r(\zb_n) $ we obtain the following result:
\begin{multline}
r(\zb_n) \propto \exp\Big(\sum_{\alpha_n\in\lk 0,1\rk}r(\alpha_n)\cdot \Big[\log p(\zb_n|\vb_n, \alpha_n) + \\ \sum_f -\log\Big({\sigma_f^{\alpha_n}(\zb_n,\vb_n)} \Big)
-\frac{|m_{fn}|^2+\nu_{fn}}{\sigma_f^{\alpha_n}(\zb_n,\vb_n)}\Big]\Big).
\label{eq:rz}
\end{multline}
The above distribution cannot be computed in closed-from. Nevertheless, we can draw samples from it using the Metropolis-Hastings (MH) algorithm \cite{bishop06}.

Let $ \tilde{r}(\zb_n) $ denote the right-hand side of \eqref{eq:rz}. At the iteration $ m $ of the MH algorithm, given a current sample $ \zb_n^{(m-1)} $, to obtain the next one, i.e., $ \zb_n^{(m)} $, we first draw a candidate sample from a proposal Gaussian distribution centered around $ \zb_n^{(m-1)} $ and with $ \epsilon\Ib $ as the covariance matrix. The candidate sample, denoted $ \zb_n^{(T)} $, is accepted by the probability $ p = \min\Big(1, \frac{\tilde{r}(\zb_n^{(T)})}{\tilde{r}(\zb_n^{(m-1)})}\Big) $. If the sample is accepted, we set $ \zb_n^{(m)}= \zb_n^{(T)}$. Otherwise, $ \zb_n^{(m)}= \zb_n^{(m-1)} $.  The above procedure is repeated until the required number of samples is achieved. Furthermore, the initial samples, obtained during the so-called burn-in period, are discarded. 
\subsection{E-$ \alpha_n $ step}
To update the variational distribution of $ \alpha_n $, we can write:
\begin{multline}
r(\alpha_n)\propto \exp \Big(\E_{r(\sbb_n)\cdot r(\zb_n)} \Big[\log p(\sbb_n|\zb_n,\vb_n,\alpha_n)+\\ \log p(\zb_n|\vb_n,\alpha_n)+\log p(\alpha_n)\Big]\Big)
\end{multline}
which is a Bernoulli distribution with
\begin{equation}
\pi_n = g\Big(\E_{r(\sbb_n)\cdot r(\zb_n)} \Big[\log\frac{p(\sbb_n,\zb_n|\vb_n,\alpha_n=1)}{p(\sbb_n,\zb_n|\vb_n,\alpha_n=0)}\Big] + \log \frac{\pi}{1-\pi}  \Big)
\end{equation}
as the parameter, where $ g(.) $ denotes the sigmoid function defined as $ g(x) = 1/(1+\exp(-x)) $. The above expression is further simplified to the following:
\begin{align}
\pi_n &\approx g\Big(\frac{1}{D}\sum_{d=1}^{D}\sum_f\log\frac{\sigma_f^{av}(\zb_n^{(d)},\vb_n)}{\sigma_f^{a}(\zb_n^{(d)})}+\label{eq:post_alpha}\\
\nonumber&\Big(1/\sigma_f^{av}(\zb_n^{(d)},\vb_n)-1/\sigma_f^{a}(\zb_n^{(d)})\Big)\cdot\Big(|m_{fn}|^2+\nu_{fn}\Big) +\\
\nonumber&\sum_l \log\bar{\sigma}_{l}(\vb_n)+\frac{(z_{ln}^{(d)}-\bar{\mu}_l(\vb_n))^2}{2\bar{\sigma}_{l}(\vb_n)}-\frac{(z_{ln}^{(d)})^2}{2}+ \log 
\frac{\pi}{1-\pi}  \Big)
\end{align}
where the expectation with respect to $ r(\zb_n) $ has been approximated by a Monte-Carlo average.
\subsection{M step}
The parameters of the mixture model, that is, $ \lk\mathbf{W}_b,\mathbf{H}_b,\pi \rk $ are estimated by optimizing the expected data log-likelihood, which takes the following form:
\begin{multline}
Q(\mathbf{W}_b,\mathbf{H}_b,\pi)=\sum_{(f,n)} -\frac{|x_{fn}-m_{fn}|^2+\nu_{fn}}{\left(\mathbf{W}_b\mathbf{H}_b\right)_{fn}}-\\\log\left(\mathbf{W}_b\mathbf{H}_b\right)_{fn}+
\pi_n\cdot\log \pi + (1-\pi_n)\cdot\log(1-\pi).
\label{eq:q_func}
\end{multline}
The update formulas for $ \mathbf{W}_b $ and $ \mathbf{H}_b $ are then obtained by using standard multiplicative update rules \cite{FevotBD09}:
\begin{equation}
\mathbf{H}_b \leftarrow \mathbf{H}_b \odot  \frac{\mathbf{W}_b^\top \left( \Vb \odot
	\left(\mathbf{W}_b \mathbf{H}_b\right)^{\odot-2} \right)}{\mathbf{W}_b^\top \left(\mathbf{W}_b \mathbf{H}_b\right)^{\odot-1} },
\label{updateH}
\end{equation}
\begin{equation}
\mathbf{W}_b \leftarrow \mathbf{W}_b \odot \frac{ \left( \Vb \odot
	\left(\mathbf{W}_b \mathbf{H}_b\right)^{\odot-2} \right)\mathbf{H}_b^\top}{\left(\mathbf{W}_b \mathbf{H}_b\right)^{\odot-1}\mathbf{H}_b^\top  },
\label{updateW}
\end{equation}
where
\begin{equation}
\Vb = \Big[|x_{fn}-m_{fn}|^2+\nu_{fn}\Big]_{(f,n)}.
\end{equation}
The prior probability of $ \alpha_n $ is also updated as follows:
\begin{equation}
\label{eq:pi-up}
\pi=\frac{1}{N}\sum_{n=1}^{N}\pi_n.
\end{equation}
\subsection{Speech enhancement}\label{subsec:se}
After the convergence of the VEM, the speech \ac{STFT} frames are estimated as their posterior means. That is, $ \forall (f,n) $:
\begin{equation}
\hat{s}_{fn} = \E_{r(s_{fn})}[s_{fn}]=\frac{\gamma_{fn}}{\gamma_{fn} + \left(\mathbf{W}_b\mathbf{H}_b\right)_{fn}}\cdot x_{fn}
\label{source_estimate}
\end{equation}
where all the involved parameters are set to their optimal values obtained in the VEM framework. 

The complete speech enhancement algorithm is summarized in Algorithm~\ref{algo:prop_alg}. Note that by 
setting $ \forall n:~\pi_n=\pi=1 $ (respectively, $ \forall n:~\pi_n=\pi=0 $), this algorithm reduces to an A-VAE (respectively, AV-VAE) based speech 
enhancement method.

\section{Experiments}\label{sec:exp}
\textit{Dataset and models.} We considered two trained VAE models: an A-VAE and an AV-VAE, from \cite{sadeghiLAGH19}, which have been trained on the NTCD-TIMIT dataset \cite{Abde17}. The test set includes about $ 1 $ hour noisy speech, along with their corresponding lips \ac{ROI}s, with six different noise types, including \textit{\ac{LR}}, \textit{White}, \textit{Cafe}, \textit{Car}, \textit{Babble}, and \textit{Street}, with noise levels: $ \lk -15,-10,-5,0,5,10 \rk $~dB. 
\begin{algorithm}[t]
	\caption{VAE-MM for speech enhancement}\label{alg:prop}
	\begin{algorithmic}[1]
		\State \textbf{Inputs:} \begin{itemize}
			\item Learned A-VAE and AV-VAE models \cite{sadeghiLAGH19}
			\item Noisy microphone STFT frames $ \mathbf{x} = \lk 
			\mathbf{x}_n\rk_{n=0}^{N-1} $
			\item Visual embeddings $ \mathbf{v} = \lk \mathbf{v}_n\rk_{n=0}^{N-1} $
		\end{itemize} 
		\State \textbf{Initialization:}
		\begin{itemize}
			\item Initialize the NMF noise parameters $ \Hb_b $ and $ \Wb_b $ with random nonnegative values, and set $\pi=0.5$.
			\item Initialize the latent codes $ \mathbf{z}^a = \lk \mathbf{z}_n^{a}\rk_{n=0}^{N-1} $ (A-VAE) and $ \mathbf{z}^{av} = \lk 
\mathbf{z}_n^{av}\rk_{n=0}^{N-1} $ (AV-VAE) using the corresponding learned encoder networks with $ \mathbf{x} $ and $ \mathbf{v} $. Then, set $ \mathbf{z} = 
\lk {\pi\cdot\mathbf{z}_n^{a}+(1-\pi)\cdot\mathbf{z}_n^{av}}\rk_{n=0}^{N-1} $.
% $= \lk \mathbf{x}_n\rk_{n=0}^{N-1} $ and $ \mathbf{v} = \lk \mathbf{v}_n\rk_{n=0}^{N-1} $ as inputs. Then, set $ \mathbf{z} = \lk 
% {\pi\cdot\mathbf{z}_n^{a}+(1-\pi)\cdot\mathbf{z}_n^{av}}\rk_{n=0}^{N-1} $.	
		\end{itemize} 	
		\While{stop criterion not met}:
		\begin{itemize}
		\item { \textbf{E-$\zb$ step:} Sample from \eqref{eq:rz} by the MH algorithm.}
		\item { \textbf{E-$\sbb$ step:}} Update $ \sbb_n $'s posterior by \eqref{eq:rs0}.
		\item { \textbf{E-$\alpha$ step:}} Update $ \alpha_n $'s posterior by \eqref{eq:post_alpha}.
		\item \textbf{M-step:} Update the parameters using \eqref{updateH} -- \eqref{eq:pi-up}.
% 		\item \textbf{M-$ \pi $-step:} Update the prior probability using \eqref{eq:pi-up}.
% 		\item  \textbf{M-$ (\Wb_b, \Hb_b) $-step:} Use \eqref{updateH} and \eqref{updateW}.
		\end{itemize}
		\EndWhile
		\State \textbf{Speech enhancement:} using \eqref{source_estimate}.
% 		Compute $ \mathbf{s} = \lk 
% 		\mathbf{s}_n\rk_{n=0}^{N-1} $ with \eqref{source_estimate}.
		
	\end{algorithmic}
	\vspace{-2pt}
	\label{algo:prop_alg}
\end{algorithm}
\begin{figure}[t]
	\centering
	\subfloat[PESQ]{{\includegraphics[width=3.85cm]{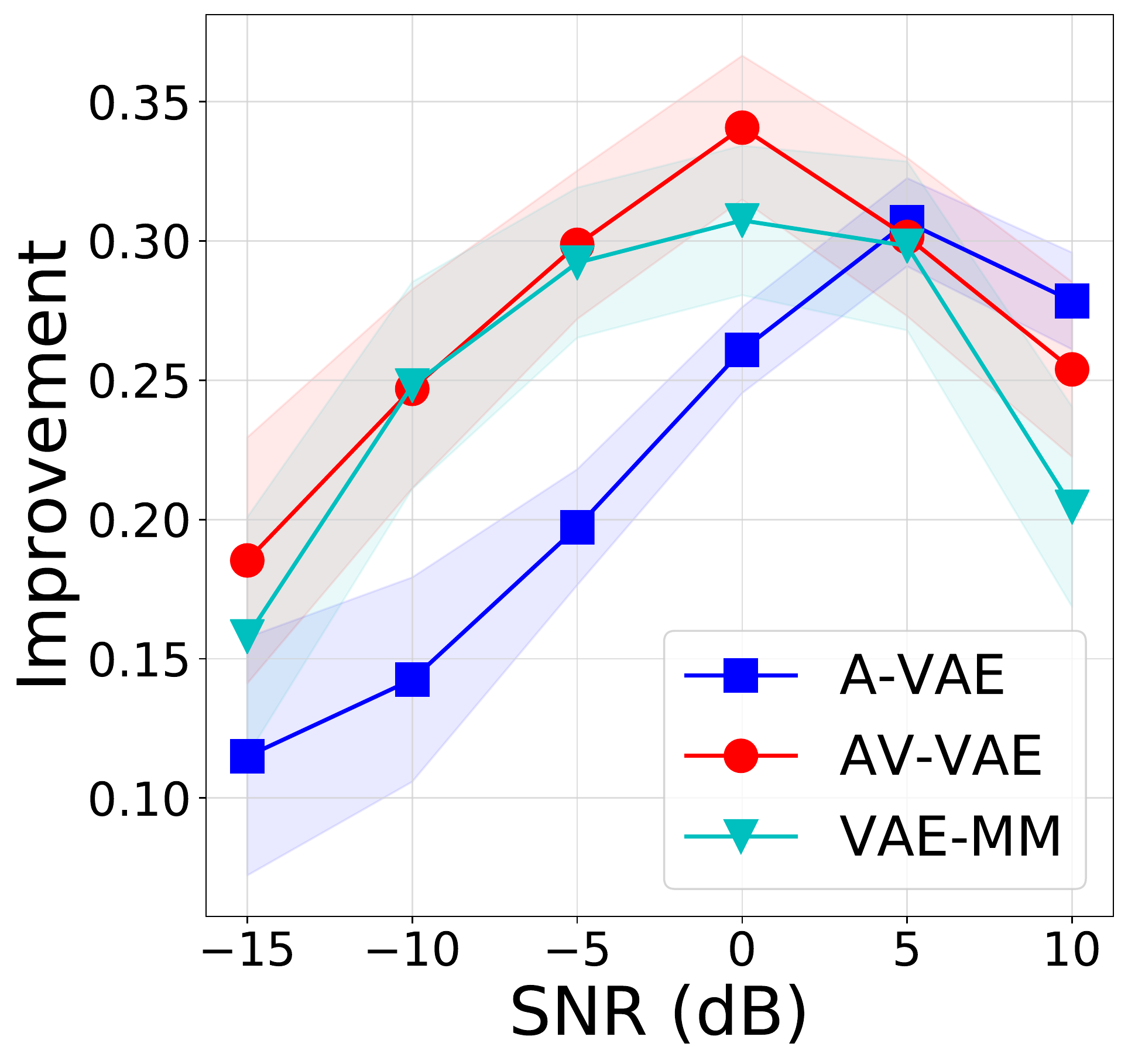} }}\vspace{-2mm}
	\subfloat[SDR]{{\includegraphics[width=3.85cm]{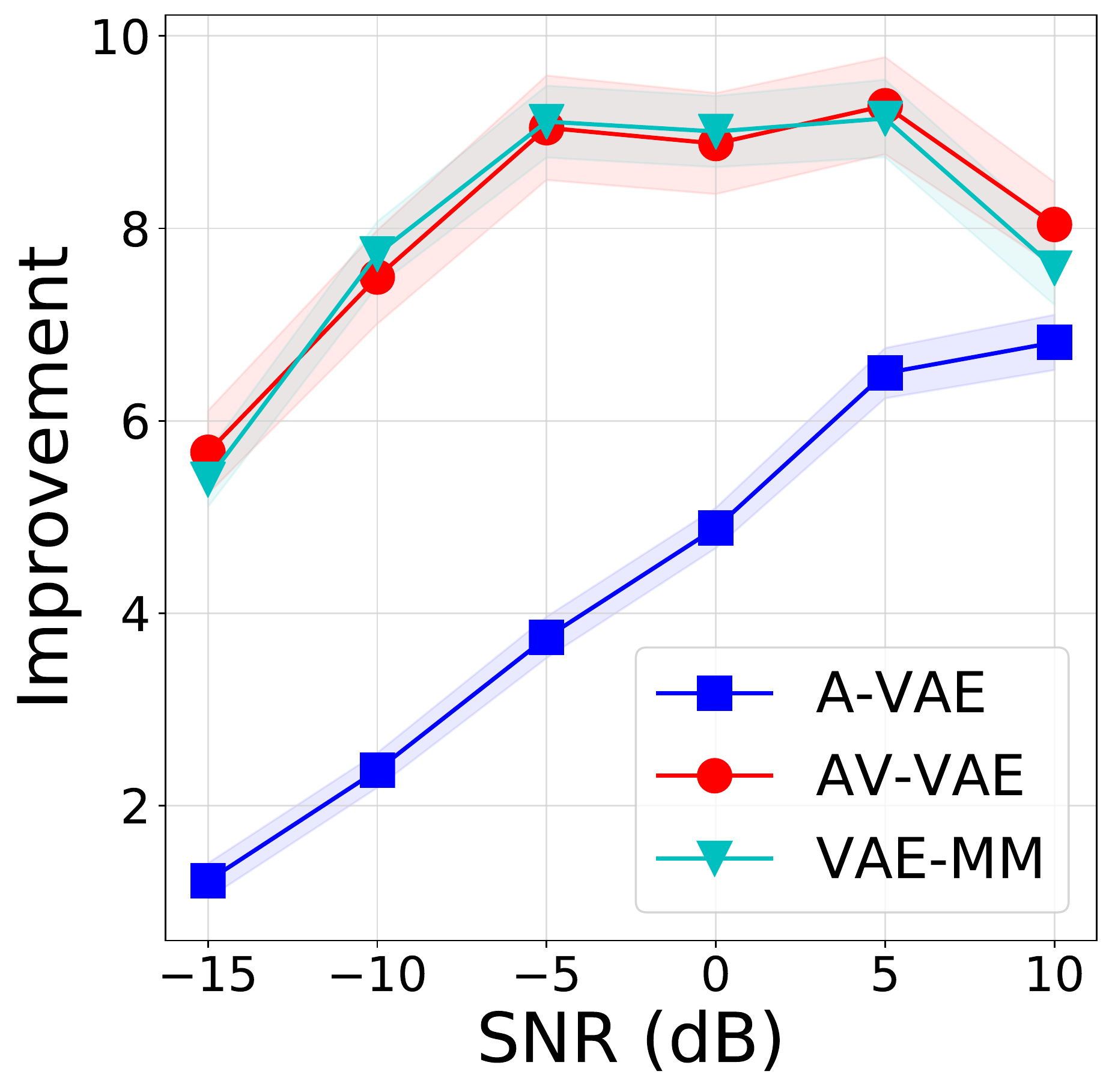} }}\\
	\subfloat[PESQ]{{\includegraphics[width=3.85cm]{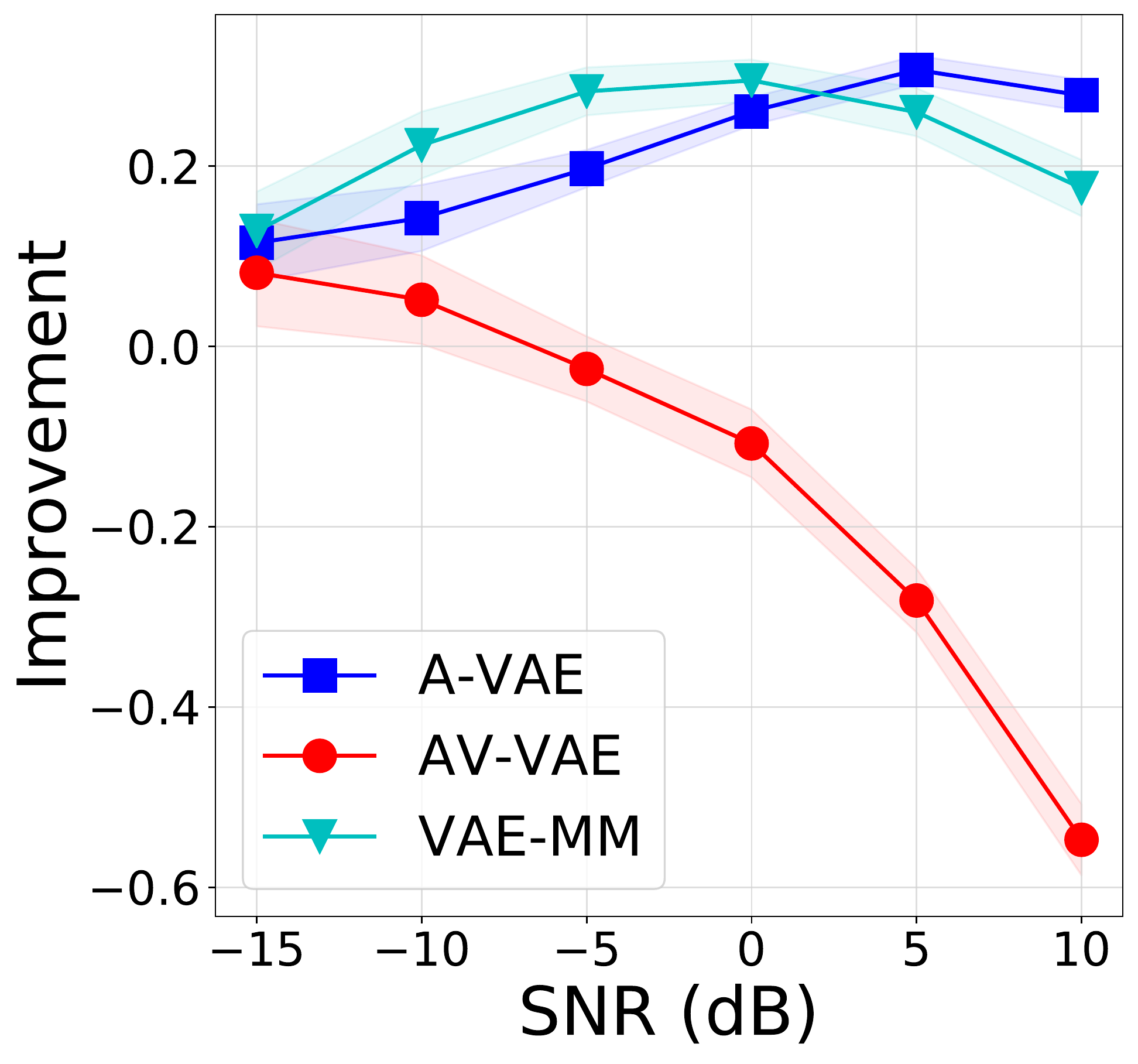} }}
	\subfloat[SDR]{{\includegraphics[width=3.75cm]{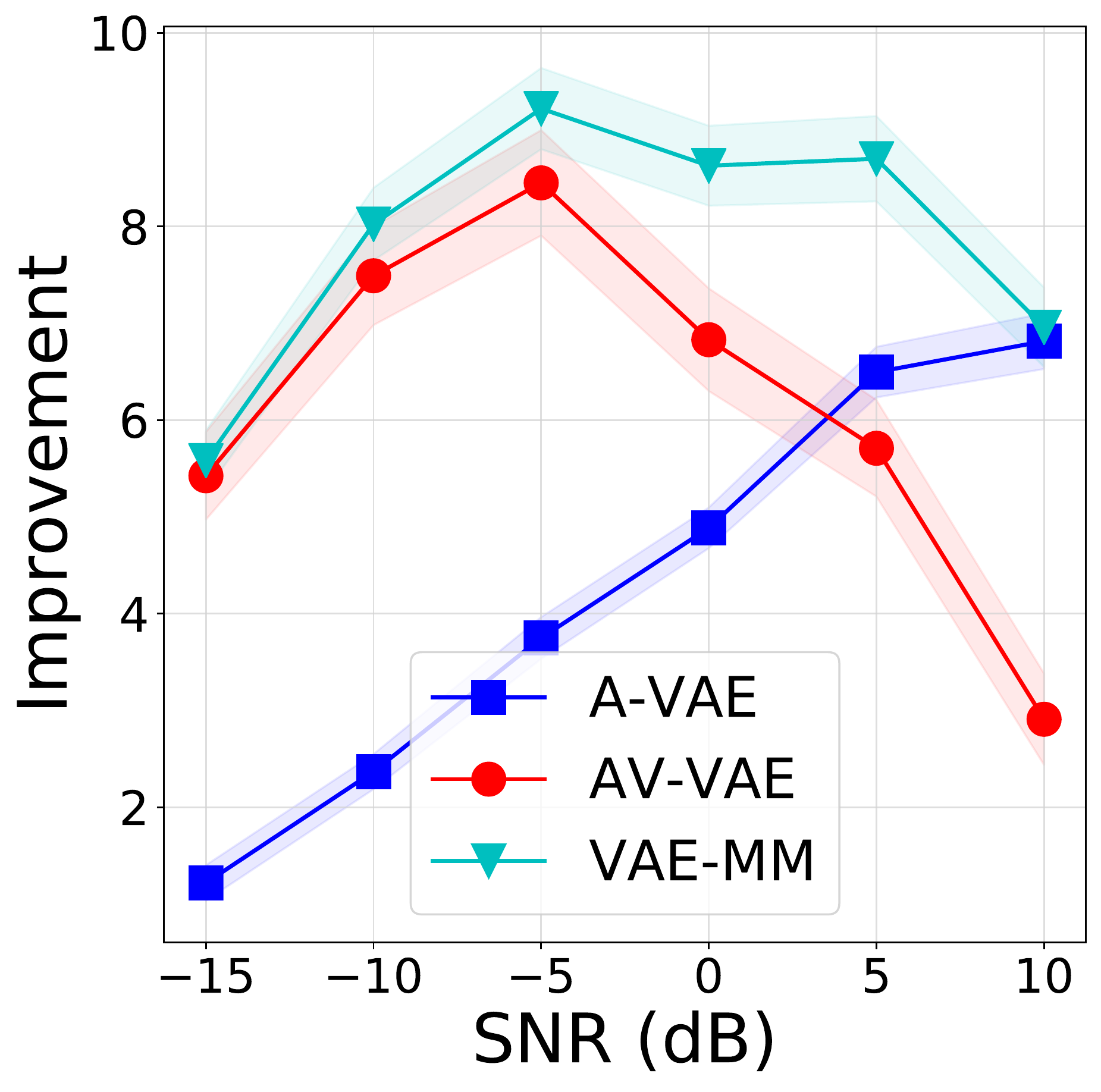} }}	
	\caption{\label{fig:mix_a_av_Fast} Speech enhancement performance for clean (top) and noisy (bottom) visual data.}
	\vspace{-3mm}
\end{figure}

\textit{Parameters settings.} For the MH
algorithm, similarly to \cite{Leglaive_MLSP18, sadeghiLAGH19}, we run 40 iterations using $\epsilon = 0.01 $ for the proposal distribution. The first 30 samples were discarded. The VEM steps were performed for $ 200 $ iterations. We initialize $ m_{fn}=x_{fn} $ and $ \nu_{fn}=0 $, $ \forall (f,n) $. The latent codes, $ \zb = \lk 
\mathbf{z}_n\rk_{n=0}^{N-1} $, were initialized as described in Algorithm~\ref{algo:prop_alg}.

\textit{Experimental protocol.} In the following, we compare the performance of A-VAE, AV-VAE, and the proposed VAE-MM. First, we run the three methods on clean 
visual data. In the second experiment, we randomly corrupt about $ 1/3 $ of the total lips images per test instance. The occluded 
images are created by randomly selecting sub-sequences of 20 consecutive video frames and adding to the associated lips images random patches of standard Gaussian noise. We used two standard speech enhancement scores, i.e., the \ac{SDR} \cite{vincent2006performance} and the \ac{PESQ}  \cite{rix2001perceptual} scores. \ac{SDR} is measured in 
decibels (dB) while \ac{PESQ} values lie in the interval $[-0.5,4.5]$ (higher the better). For computing SDR, the mir\_eval Python library was 
used.\footnote{\url{https://github.com/craffel/mir_eval}} For each measure, we report the difference between the output value, i.e., evaluated on the enhanced 
speech signal, and the input value, i.e., evaluated on the noisy mixture. 

\textit{Results.} Figure~\ref{fig:mix_a_av_Fast} summarizes the results. First, as can be seen, AV-VAE shows a much better performance than A-VAE when visual data is clean. Second, in the case of the clean visual data, VAE-MM and AV-VAE show similar performances in terms of SDR. In the PESQ measure, we see some small drops in the performance of VAE-MM compared to that of AV-VAE. For noisy visual data, AV-VAE's performance drops significantly. However, VAE-MM seems to have successfully skipped the occluded lips images by switching to A-VAE. Its performance is still better than that of A-VAE, as some of the video frames contain clean and usable visual data.

\section{Conclusion}
In this paper, we proposed an audio-visual speech generative model based on a VAE mixture consisting of a trained A-VAE and a trained AV-VAE. Combined with an NMF model for noise variance \cite{bando2018statistical, Leglaive_MLSP18, sadeghiLAGH19}, the goal was to make audio-visual VAE speech enhancement robust against noisy visual frames by switching to A-VAE in an unsupervised way. We presented a variational 
expectation-maximization approach to estimate the parameters of the model as well as the clean speech. The promising performance of the proposed method was 
demonstrated through some experiments.

% The 
% main idea is to combine an audio-only VAE (A-VAE) and an AV-VAE within a probabilistic mixture model. The parameters of this model, including membership 
% probabilities and also the non-negative matrices modeling noise variance, are estimated using a variational expectation-maximization approach. Through a number 
% of experiments with clean and occluded video frames, the promising performance of the proposed method was confirmed.
\balance
\bibliographystyle{IEEEbib} %IEEEbib
\bibliography{myref_compressed}

\begin{thebibliography}{10}

\bibitem{VincVG18}
E.~Vincent, T.~Virtanen, and S.~Gannot,
\newblock {\em Audio Source Separation and Speech Enhancement},
\newblock Wiley, 2018.

\bibitem{wang2018supervised}
W.~DeLiang and J.~Chen,
\newblock ``Supervised speech separation based on deep learning: An overview,''
\newblock {\em IEEE Transactions on Audio, Speech, and Language Processing},
  vol. 26, no. 10, pp. 1702--1726, 2018.

\bibitem{loizou2007speech}
P.~C. Loizou,
\newblock {\em Speech enhancement: theory and practice},
\newblock CRC press, 2007.

\bibitem{xu2015regression}
Yong Xu, Jun Du, Li-Rong Dai, and Chin-Hui Lee,
\newblock ``A regression approach to speech enhancement based on deep neural
  networks,''
\newblock {\em IEEE Transactions on Audio, Speech, and Language Processing},
  vol. 23, no. 1, pp. 7--19, 2015.

\bibitem{bando2018statistical}
Y.~Bando, M.~Mimura, K.~Itoyama, K.~Yoshii, and T.~Kawahara,
\newblock ``Statistical speech enhancement based on probabilistic integration
  of variational autoencoder and non-negative matrix factorization,''
\newblock in {\em Proc. IEEE International Conference on Acoustics, Speech, and
  Signal Processing (ICASSP)}, 2018, pp. 716--720.

\bibitem{Leglaive_MLSP18}
S.~Leglaive, L.~Girin, and R.~Horaud,
\newblock ``A variance modeling framework based on variational autoencoders for
  speech enhancement,''
\newblock in {\em Proc. IEEE International Workshop on Machine Learning for
  Signal Processing (MLSP)}, 2018, pp. 1--6.

\bibitem{SekiguchiAPSIPA2018}
K.~Sekiguchi, Y.~Bando, K.~Yoshii, and T.~Kawahara,
\newblock ``Bayesian multichannel speech enhancement with a deep speech
  prior,''
\newblock in {\em Proc. Asia-Pacific Signal and Information Processing
  Association Annual Summit and Conference (APSIPA ASC)}, 2018, pp. 1233--1239.

\bibitem{Leglaive_ICASSP2019a}
S.~Leglaive, L.~Girin, and R.~Horaud,
\newblock ``Semi-supervised multichannel speech enhancement with variational
  autoencoders and non-negative matrix factorization,''
\newblock in {\em Proc. IEEE International Conference on Acoustics, Speech, and
  Signal Processing (ICASSP)}, 2019, pp. 101--105.

\bibitem{Leglaive_ICASSP2019b}
S.~Leglaive, U.~\c{S}im\c{s}ekli, A.~Liutkus, L.~Girin, and R.~Horaud,
\newblock ``Speech enhancement with variational autoencoders and alpha-stable
  distributions,''
\newblock in {\em Proc. IEEE International Conference on Acoustics, Speech, and
  Signal Processing (ICASSP)}, 2019, pp. 541--545.

\bibitem{PariDV19}
M.~Pariente, A.~Deleforge, and E.~Vincent,
\newblock ``A statistically principled and computationally efficient approach
  to speech enhancement using variational autoencoders,''
\newblock in {\em Proc. Conference of the International Speech Communication
  Association (INTERSPEECH)}, 2019.

\bibitem{girin2001audio}
L.~Girin, J.-L. Schwartz, and G.~Feng,
\newblock ``Audio-visual enhancement of speech in noise,''
\newblock {\em The Journal of the Acoustical Society of America}, vol. 109, no.
  6, pp. 3007--3020, 2001.

\bibitem{AfouCZ18}
T.~Afouras, J.~S. Chung, and A.~Zisserman,
\newblock ``The conversation: {D}eep audio-visual speech enhancement,''
\newblock in {\em Proc. Conference of the International Speech Communication
  Association (INTERSPEECH)}, 2018, pp. 3244--3248.

\bibitem{GabbSP18}
A.~Gabbay, A.~Shamir, and S.~Peleg,
\newblock ``Visual speech enhancement,''
\newblock in {\em Proc. Conference of the International Speech Communication
  Association (INTERSPEECH)}, 2018, pp. 1170--1174.

\bibitem{sadeghiLAGH19}
M.~Sadeghi, S.~Leglaive, X.~Alameda-Pineda, L.~Girin, and R.~Horaud,
\newblock ``Audio-visual speech enhancement using conditional variational
  auto-encoder,''
\newblock {\em arXiv preprint arxiv.org/abs/1908.02590}, 2019.

\bibitem{KingW14}
D.~P. Kingma and M.~Welling,
\newblock ``Auto-encoding variational bayes,''
\newblock in {\em International Conference on Learning Representations (ICLR)},
  2014.

\bibitem{bishop06}
C.~Bishop,
\newblock {\em Pattern Recognition and Machine Learning},
\newblock Springer-Verlag Berlin, Heidelberg, 2006.

\bibitem{FevotBD09}
C.~F{\'e}votte, N.~Bertin, and J.-L. Durrieu,
\newblock ``{Nonnegative matrix factorization with the Itakura-Saito
  divergence: With application to music analysis},''
\newblock {\em Neural computation}, vol. 21, no. 3, pp. 793--830, 2009.

\bibitem{Abde17}
A.-H. Abdelaziz,
\newblock ``{NTCD-TIMIT}: A new database and baseline for noise-robust
  audio-visual speech recognition,''
\newblock in {\em Proc. Conference of the International Speech Communication
  Association (INTERSPEECH)}, 2017, pp. 3752--3756.

\bibitem{vincent2006performance}
E.~Vincent, R.~Gribonval, and C.~F{\'e}votte,
\newblock ``Performance measurement in blind audio source separation,''
\newblock {\em IEEE Transactions on Audio, Speech, and Language Processing},
  vol. 14, no. 4, pp. 1462--1469, 2006.

\bibitem{rix2001perceptual}
A.~W. Rix, J.~G. Beerends, M.~P. Hollier, and A.~P. Hekstra,
\newblock ``{Perceptual evaluation of speech quality (PESQ)-a new method for
  speech quality assessment of telephone networks and codecs},''
\newblock in {\em Proc. IEEE International Conference on Acoustics, Speech, and
  Signal Processing (ICASSP)}, 2001, pp. 749--752.

\end{thebibliography}

\end{document}